\documentclass[conference]{IEEEtran}
\IEEEoverridecommandlockouts
\usepackage{cite}
\usepackage{makecell}
\usepackage{amsmath,amssymb,amsfonts}
\usepackage{graphicx}
\usepackage{xcolor}
\usepackage{stfloats}
\usepackage{tcolorbox}
\usepackage{multirow}
\usepackage{subcaption}
\usepackage{array}
\usepackage{textcomp}
\usepackage{url}
\usepackage{verbatim}
\usepackage{booktabs}
\def\BibTeX{{\rm B\kern-.05em{\sc i\kern-.025em b}\kern-.08em
    T\kern-.1667em\lower.7ex\hbox{E}\kern-.125emX}}
\def\BibTeX{{\rm B\kern-.05em{\sc i\kern-.025em b}\kern-.08em
    T\kern-.1667em\lower.7ex\hbox{E}\kern-.125emX}}

\usepackage{balance}
\usepackage[normalem]{ulem}
\begin{document}

\title{Impact of PA Nonlinearities on AFDM Sensing: A Matched Filtering Perspective}
\author{\IEEEauthorblockA{Eya Gourar\IEEEauthorrefmark{1}, Yahia Medjahdi\IEEEauthorrefmark{1}, Laurent Clavier\IEEEauthorrefmark{1}\IEEEauthorrefmark{3}, Abdul Karim Gizzini\IEEEauthorrefmark{2}, Patrick Sondi\IEEEauthorrefmark{1}
}\\
\IEEEauthorblockA{\IEEEauthorrefmark{1} IMT Nord Europe, Institut Mines T\'el\'ecom, Center for Digital Systems, F-59653 Villeneuve d’Ascq, France\\
\IEEEauthorrefmark{3} Inria, Villeneuve-d’Ascq, France\\
\IEEEauthorrefmark{2} University of Paris-Est Créteil (UPEC), LISSI/TincNET, F-94400, Vitry-sur-Seine, France\\
Email: \ eya.gourar@imt-nord-europe.fr}
}
\maketitle


\begin{abstract}
Affine Frequency Division Multiplexing (AFDM) is a promising waveform for integrated sensing and communications (ISAC) due to its Doppler resilience. However, such waveforms exhibit a high peak-to-average power ratio (PAPR), and they require highly linear Power Amplifiers (PAs), which conflicts with ISAC requirements, where high transmit power for sensing is always beneficial. This paper investigates the impact of PA nonlinearities on the sensing performance of AFDM. Using the Bussgang decomposition of the nonlinearly amplified AFDM signal, we analyze the delay-Doppler ambiguity function for the sensing matched filtering and show that AFDM’s sidelobes remain nearly unaffected. Simulation results confirm that AFDM outperforms OFDM in target detection under high Doppler and nonlinear amplification, highlighting its suitability for the sensing task.
\end{abstract}


\section{Introduction}
Affine frequency division multiplexing (AFDM) has emerged as a promising alternative to orthogonal frequency division multiplexing (OFDM) for sixth-generation (6G) communications, as it better matches the technical requirements of future use cases than OFDM \cite{chafii2023twelve}. In particular, it is more robust against highly time- and frequency-selective propagation channels, as they can achieve full diversity in such environments, unlike OFDM \cite{bemani2023affine,rou2024otfs}. Beyond communication performance, AFDM has also attracted considerable attention in the context of Integrated Sensing and Communication (ISAC), which has been identified as one of the key usage scenarios for future 6G networks. 
The sensing capability of a waveform is typically evaluated using the Ambiguity Function (AF), which characterizes the delay-Doppler response of the sensing Matched Filter (MF) at the receiver. In particular, the AF's sidelobes designate the level of interference from adjacent targets, a factor that is crucial for multi-target sensing in 6G.

Recent studies (e.g, \cite{liu2025cp,zhang2025discrete}) show that among orthogonal waveforms, OFDM achieves the lowest ranging sidelobe levels, leading to high ranging accuracy. More recently, several works have investigated the AF and sensing performance of AFDM waveforms \cite{ni2022afdm,bemani2024integrated,yin2025ambiguity,zhang2025discrete,rou2025normalized,ni2025ambiguity,bedeer2025ambiguity,zhu2024afdm,zhang2025afdm}. The merits of AFDM-based sensing using the $\mathrm{DAFT}$-domain matched filtering were demonstrated in \cite{ni2022afdm}. \cite{zhu2024afdm} provided a preliminary analysis of the AF behavior and corresponding chirp-parameter design rule. In \cite{yin2025ambiguity}, the authors provide a rigorous characterization of the continuous-time AF and develop the impact of chirp parameters on sensing resolution. A system-level ISAC framework with pilot-assisted sensing design and AF formulation was established in \cite{zhang2025afdm}. Extensions of the AF to pulse-shaped and random signaling were studied in \cite{ni2025ambiguity}, while comparative normalized AF analyses across OFDM-, Orthogonal Time Frequency and Space (OTFS)-, and AFDM-based waveforms were reported in \cite{rou2025normalized}. More generally, discrete ambiguity properties of random communication waveforms were analyzed in \cite{zhang2025discrete}, providing theoretical tools applicable to AFDM-based sensing.

However, despite recent advances, several important aspects remain insufficiently explored. In particular, practical transmitter impairments such as nonlinear power amplifiers (PAs) can significantly distort the transmission waveforms. For sensing, PAs are intentionally operated close to saturation to maximize radiated power and coverage, in order to overcome the two-way path loss of the radar channel, making the presence of nonlinearities unavoidable. And, while operating at a smaller amplification level can reduce these distortions, it severely degrades the transmitter's power efficiency. The impact of PA nonlinearities on the sensing task and the AF of OFDM has been extensively studied (see e.g \cite{gourar2025ambiguity,ismail2024robustness}). However, the robustness of AFDM's sensing to such distortions has not yet been investigated. Consequently, it remains unclear whether the conclusions established for OFDM also hold for AFDM-based systems. Therefore, in this paper, we aim to explore this area. The main contributions are summarized as follows:
\begin{itemize}
    \item We analyze the AF of a nonlinearly amplified AFDM via the Bussgang decomposition, and we provide further characterizations of the zero-Doppler and zero-delay cuts.
    \item We compare AFDM and OFDM in terms of the average AF's Peak Sidelobe Level Ratio (PSLR), and target detection performance under PA nonlinearities and high-Doppler conditions. In addition, we suggest analyzing the complementary cumulative distribution function (CCDF) of AF-based metrics to characterize the outage probability of unfavorable sidelobe behaviour under nonlinearities.
    \item We show that AFDM demonstrates the capability to reliably detect targets in scenarios where OFDM fails, characterized by severe PA nonlinearities and Doppler effects.
\end{itemize}
The paper is organized as follows: Section \ref{section1} describes the system model. Section \ref{section2} analyzes the Ambiguity function under PA nonlinearities. Finally, section \ref{section3} presents the simulation results, prior to our conclusions.

\section{System Model}\label{section1}

We present an AFDM-based system, where the base station transmits the AFDM waveform to the downlink users and simultaneously receives echoes reflected by the targets around it.
\subsection{AFDM Signal Model} 
In AFDM, a one-dimensional vector $\mathbf{s}\in \mathbb{C}^{N \times 1}$ is multiplexed into the twisted time-frequency chirp domain using the inverse discrete affine Fourier transform ($\mathrm{IDAFT}$) \cite{bemani2021afdm},
\begin{align}
\mathbf{x}= \mathbf{A}^H  \mathbf{s}
 = (\mathbf{\Lambda}_{c_1}^{H}  \mathbf{F}^{H}  \mathbf{\Lambda}_{c_2}^{H})  \;\mathbf{s}\in \mathbb{C}^{N \times 1},
\label{afdm_idaft}
\end{align}
where $(\cdot)^H$ is the conjugate transpose, $\mathbf{A} = \mathbf{\Lambda}_{c_2} \mathbf{F} \mathbf{\Lambda}_{c_1} \in \mathbb{C}^{N \times N}$ is the forward $N$-point discrete $\mathrm{AFT}$ ($\mathrm{DAFT}$) matrix, and $\mathbf{\Lambda}_{c_i} = \mathrm{diag}\big[e^{-j 2 \pi c_i 0^2}, \dots, e^{-j 2 \pi c_i (N-1)^2}\big] \in \mathbb{C}^{N \times N}$ is a diagonal chirp matrix with central digital frequency $c_i$. $\mathbf{F} \in \mathbb{C}^{N\times N}$ is the $N$-point discrete FT ($\mathrm{DFT}$) matrix, and hence, $\mathbf{F}^H$ is the $N$-point inverse $\mathrm{DFT}$ ($\mathrm{IDFT}$) matrix. 
To combat inter-symbol interference between multiple received symbol copies due to multipath effects, a chirp-periodic prefix (CPP) of length $L$ is appended \cite{ni2022afdm}.

\subsection{Power amplifier nonlinearities modeling}
Due to PA nonlinearity at the transmitter, the amplified signal is subject to distortion. Without loss of generality, we approximate the output signal using the Bussgang theorem. For sufficiently large block lengths $N$, the central limit theorem (CLT) implies that the TD signal converges in distribution to a circularly symmetric complex Gaussian random variable with variance $\sigma_x^2$. Hence, the output of the nonlinear function, \textit{i.e.}, nonlinear PA, can be expressed according to the Bussgang approximation as \cite{dardari2002theoretical},
\vspace{-0.1cm}
\begin{equation}
\mathbf{y}= \kappa \mathbf{x}+ \mathbf{d}=\kappa\mathbf{A}^H\mathbf{s}+\mathbf{A}^H\mathbf{t}
\label{eq:y}
\end{equation}
where $\mathbf{d}$ is the PA-induced distortion term uncorrelated with $\mathbf{x}$, and with variance $\mathbf{\sigma}^2_d = \mathbb{E}[| \mathbf{d}|^2]$. Its $\mathrm{DAFT}$-domain representation is $\mathbf{t}=\mathbf{A}\mathbf{d}$. Bussgang gain $\kappa$ is defined as,
\begin{equation}
\kappa = \frac{\mathbb{E}[\mathbf{x}^{\ast}\mathbf{y}]}{\mathbb{E}[|\mathbf{x}|^2]}. \label{kappa}
\end{equation}
\subsection{Received Signal Model}
Assuming a sparse multipath environment with $P$ reflectors, the received signal can be expressed after CPP removal as,
\vspace{-0.1cm}
\begin{equation}
\mathbf{r}=\mathbf{H}\mathbf{y}+\mathbf{w} \in \mathbb{C}^{N \times 1},
\label{eq:r}
\end{equation}
where $\mathbf{w} \sim \mathcal{CN}(\mathbf{0}, \sigma_w^2 \mathbf{I}_N)$ is the additive white Gaussian noise, and $\mathbf{H}$ is the discrete channel matrix given by \cite{rou2024otfs},
\vspace{-0.2cm}
\begin{equation}
\mathbf{H}=
\sum_{i=1}^{P}h_i\mathbf{\Delta}_{k_i}\mathbf{\Pi}_{l_i} \in \mathbb{C}^{N \times N}.
\end{equation}

The complex path gain, the delay and Doppler indices are denoted as $h_i$, $l_i$ and $k_i$, respectively. $\mathbf{\Pi}_{l_i} \in \mathbb{C}^{N \times N}$ is a circular permutation matrix corresponding to delay $l_i$, and $\mathbf{\Delta}_{k_i}=\mathrm{diag}\!\left(e^{-j2\pi k_i n/N}\right)$ is the Doppler modulation matrix \cite{rou2024otfs}.
Substituting \eqref{eq:y} in \eqref{eq:r}, the received signal becomes
\begin{align} \label{r}
\mathbf{r}
&=\kappa \mathbf{H}\mathbf{x}+\mathbf{H}\mathbf{d}+\mathbf{w} =\kappa\mathbf{H}\mathbf{A}^H\mathbf{s}+\mathbf{H}\mathbf{A}^H\mathbf{t}+\mathbf{w} .
\end{align}
For sensing, to maximize the SNR after \eqref{r}, the receiver usually processes the received signal with a matched filter (MF) first, where delays and Doppler frequency shifts will be observed. For AFDM, matched filtering in the $\mathrm{DAFT}$ domain is applied to estimate the delay-Doppler parameters for target range and velocity detection, by exploiting the AFDM-induced cyclic shifts and 2D linear phase rotations along propagation paths \cite{ni2022afdm}.

\subsection{DAFT-Domain Parameter Estimation for AFDM} \label{sec:2D-construction}

For a transmitted frame of $M$ consecutive AFDM symbols, we define the transmitted symbol matrix $
\mathbf{S} = [\mathbf{s}_0, \mathbf{s}_1, \dots, \mathbf{s}_{M-1}] \in \mathbb{C}^{N \times M}$, and the corresponding TD signals $\mathbf{X} = [\mathbf{x}_0, \mathbf{x}_1, \dots, \mathbf{x}_{M-1}] \in \mathbb{C}^{N \times M}$, where each column $\mathbf{s}_m$ contains the $N$ data symbols of the $m^\text{th}$ AFDM symbol, and $\mathbf{x}_m = \mathbf{A}^H \mathbf{s}_m$ is the corresponding TD transmitted signal as in \eqref{afdm_idaft}.

At the receiver side, the $\mathrm{DAFT}$ domain frame can be written as,
\begin{equation}
\mathbf{G} = \mathbf{A} \mathbf{R} = \kappa \mathbf{A} \mathbf{H} \mathbf{A}^H \mathbf{S} + \tilde{\mathbf{D}} + \tilde{\mathbf{W}} \in \mathbb{C}^{N \times M},
\end{equation}
where $\tilde{\mathbf{D}} = \mathbf{A} \mathbf{H} \mathbf{D}$, $\tilde{\mathbf{W}} = \mathbf{A} \mathbf{W},$ are the PA-induced distortion and noise representations in the DAFT domain, and $\mathbf{R}, \mathbf{D}, \mathbf{W} \in \mathbb{C}^{N \times M}$ are the received signal, distortion, and noise frames, respectively.

\paragraph{Compensation for Delay-Induced Linear Phase}
To remove the linear phase shift caused by the propagation delay $l_i$ along the
rows, for each candidate delay $l \in \{0,\dots,L-1\}$, we generate a compensation matrix as $\mathbf{L}_l = \mathrm{diag}\left(e^{j \frac{2\pi}{N} l p}\right)_{p=0}^{N-1}$, and apply it to the received DAFT-domain matrix as $\mathbf{Z}_l = \mathbf{L}_l \mathbf{G}$ \cite{ni2022afdm}.
When $l = l_i$, the delay-induced linear phase shift is removed, leaving only the Doppler-induced phase rotation.

\paragraph{ Matched Filtering in the $\mathrm{DAFT}$ Domain}
For each candidate delay $l\in [0,\cdots, L-1]$, the delay-compensated matrix $\mathbf{Z}_l$ is correlated with the transmitted symbols $\mathbf{S}$ using a $\mathrm{DAFT}$-domain MF \cite{ni2022afdm},
\begin{equation} \label{W_l}
\mathbf{\mathcal{A}}_l = \mathbf{F}^H \Big( (\mathbf{F}\mathbf{Z}_l)^* \odot (\kappa \mathbf{F}\mathbf{S}) \Big) \mathbf{F},
\end{equation}
where $(\cdot)^*$ denotes conjugation, and $\odot$ represents element-wise multiplication. This yields a 2D range-Doppler (RD) map in which peaks correspond to target responses.

\paragraph{ Extraction of Delay and Integer Doppler}
When $l = l_i$, the resulting matrix $\mathbf{\mathcal{A}}_l$ will contain one or more peaks, depending on how many reflectors have this delay and a different Doppler shift. The presence of a target is detected if the peak magnitude of $\mathbf{\mathcal{A}}_l$ exceeds a Constant false alarm rate (CFAR) threshold, which accounts for both noise and PA distortion.

For an estimated delay $\hat{l}_i$, the index of row of the peak exceeding the threshold in $\mathbf{\mathcal{A}}_l$, noted as $\tilde{p}_i$, provides the integer part of the normalized Doppler, respectively as \cite{ni2022afdm},
\begin{equation}
\hat{\alpha}_i = 2 N c_1 \hat{l}_i - (1 - \tilde{p}_i)_N.
\end{equation}
To anticipate mismatches between the AFDM waveform and the MF under different delay and Doppler-shift conditions, the ambiguity function (AF) is analyzed.

\section{Ambiguity function analysis} \label{section2}
The per-symbol analysis suffices to capture the waveform’s delay-Doppler representation. Therefore, the AF is calculated for one AFDM symbol as \cite{bedeer2025ambiguity}, 
\begin{equation}
 \chi(l,k) = \mathbf{s}^H\mathbf{B}_{(l,k)}\mathbf{s}, 
\, \, l=0,\dots,N-1,
\label{AF}
\end{equation}
with $l$ and $k$ denoting the normalized delay and Doppler frequency with respect to the sampling period and the sampling frequency, respectively, and $\mathbf{B}_{(l,k)}$ is the delay-Doppler response of the modulation matrix expressed as,
\begin{equation}
\mathbf{B}_{(l,k)}=\mathbf{A}\mathbf{J}_l \mathbf{D}_k\mathbf{A}^H,\quad l=0,\dots,N-1,
\label{B_lk}
\end{equation}
with $\mathbf{J}_l$ and $\mathbf{D}_k$ representing delay and Doppler phase operators in discrete form.

In practice, the signal can be subject to PA-induced distortions before transmission. To characterize their impact analytically, we extend the example of the AF in \eqref{AF} to the case of non-ideally amplified signals following the Bussgang model in \eqref{eq:y} as follows, 
\begin{align}
\chi_y(l,k)
&= |\kappa|^2\chi_{x}(l,k)
+\kappa\chi_{x,d}(l,k)
+\kappa^{\ast}\,\chi_{d,x}(l,k) \nonumber \\
& \quad+\chi_{d}(l,k). 
\label{chi_y}
\end{align}

\noindent where the self-AFs and cross-AFs are defined, respectively, as
\begin{align*}
    \chi_{u}(l,k) =  (\mathbf{A}\mathbf{u})^H\mathbf{B}_{(l,k)}(\mathbf{A}\mathbf{u}), \\
    \chi_{u,v}(l,k) =  (\mathbf{A}\mathbf{v})^H\mathbf{B}_{(l,k)}(\mathbf{A}\mathbf{u}).
\end{align*}

\noindent \textit{Remark:} 
In the 2D RD map $\mathbf{\mathcal{A}}_l$ in \eqref{W_l}, the peaks occur at indices $(\tilde{p}_i, \tilde{m}_i)$. Using the AF definition, we can interpret each element of $\mathbf{\mathcal{A}}_{\hat l_i}$ as a scaled AF sample, \textit{i.e.},
\begin{equation}
\mathbf{\mathcal{A}}_l(p,m) \approx \frac{\kappa h_i}{N} \chi(p+\hat l_i, k+\tilde{p}_i) + \eta(p,m),
\end{equation}
where $\eta(p,m)$ is the noise element at the $(p,k)^\text{th}$ delay-Doppler bin, after the MF. Hence, the 2D RD map obtained via $\mathrm{DAFT}$-domain MF is equivalent to sampling the AF of the transmitted AFDM waveform along the discrete grid defined by $N$ (the AFDM symbol length) and $M$ (number of columns in the $\mathrm{DAFT}$-domain representation).

The presence of thermal noise and PA distortion affects the amplitude and sidelobe floor of the RD map \cite{ismail2024robustness,gourar2025ambiguity}. Thus, to evaluate the sidelobe behavior after nonlinear power amplification, we use the Peak Sidelobe Level Ratio (PSLR) as the AF performance metric. It is define as the ratio of the highest sidelobe magnitude and the mainlobe peak magnitude,
\begin{equation}
\mathrm{PSLR}_{\mathrm{dB}}(l,k)
= 20 \log{10}\left(max_{\mathcal{W}_{(l,k)}} |\chi(l,k)| / |\chi(0,0)| \right)
\end{equation}
where $\mathcal{W}_{(l,k)}$ denotes the 2D sidelobe region excluding the mainlobe region. The AF of the signal, that is, the ambiguity of distance and velocity, can be individually analyzed by separating the 2D plane into the zero-Doppler and the zero-delay cuts, by setting $k=0$ and $l=0$, respectively. The delay $\mathrm{PSLR}$ and Doppler $\mathrm{PSLR}$ are obtained accordingly.

\subsection{Zero-Doppler Cut}
The zero-Doppler cut can also be regarded as the autocorrelation function of the TD signal. In this section, we study the autocorrelation properties of the AFDM in the presence of PA-induced nonlinearities. The zero-Doppler cut can be expressed from \eqref{chi_y} as
\begin{align}
\chi_y(l,0)
&=  |\kappa|^2\chi_x(l,0) + \kappa\chi_{x,d}(l,0) +\kappa^{\ast}\chi_{d,x}(l,0) +\chi_d(l,0),
\label{0-Doppler_y}
\end{align}
where $\chi_x(l,0)$ denotes the autocorrelation of the input signal, $\chi_d(l,0)$ that of the distortion, and $\kappa \chi_{x,d}(l,0) + \kappa^\ast \chi_{d,x}(0,l)$ is the cross-correlation term.

Under the Bussgang assumptions, the distortion component $\mathbf{d}$ is uncorrelated with the input signal $\mathbf{x}$, \textit{i.e.}, $\mathbb{E}[\mathbf{x}\mathbf{d}^H]=\mathbf{0}$. Therefore, the expectation of cross-terms $\mathbb{E}[\chi_{x,d}(l,0)] = \mathbb{E}[\mathbf{x}^H \mathbf{J}_l \mathbf{d}] = 0$ and  $\mathbb{E}[\chi_{d,x}(l,0)]= \mathbb{E}[\mathbf{d}^H \mathbf{J}_l \mathbf{x}] = 0$. We obtain,
\begin{equation}
\mathbb{E}[\chi_y(l,0)]
=
|\kappa|^2 \mathbb{E}[\chi_x(l,0)]
+
\mathbb{E}[\chi_d(l,0)].
\end{equation}
For AFDM, the $\mathrm{IDAFT}$ modulation matrix does not diagonalize the time-shift matrix $\mathbf{J}_l$ for non-zero $c_1$ \cite{zhang2025discrete}. Consequently, the intrinsic delay sidelobes $\chi_x(l,0)$ remain dominant, unlike in OFDM, where the $\mathrm{DFT}$ diagonalizes $\mathbf{J}_l$ and suppresses delay sidelobes. Although the distortion term $\chi_d(l,0)$ may introduce additional peaks, their impact is negligible compared to the elevated sidelobe floor already caused by the non-diagonal structure of $\mathbf{B}_{(l,0)}$. This is consistent with our findings in \cite{gourar2025ambiguity}, where QAM-based OFDM was shown to be relatively insensitive to PA nonlinearities. Since AFDM exhibits even higher intrinsic sidelobes than OFDM \cite{zhang2025discrete}, the effect of PA-induced distortion is expected to be even less significant.

\subsection{Zero-delay Cut}
We now characterize the impact of nonlinear PA distortion on the zero-delay cut using the Bussgang decomposition. 
By fixing $l=0$ in \eqref{chi_y}, the zero-delay cut of the amplified signal is expressed as
\vspace{-0.1cm}
\begin{align}
\chi_y(0,k) 
&= |\kappa|^2 \chi_x(0,k) + \Delta_{\text{Doppler}},
\label{chi_y_doppler}
\end{align}
where $\Delta_{\text{Doppler}} = \kappa\chi_{x,d}(0,k) +\kappa^{\ast}\chi_{d,x}(0,k) +\chi_d(0,k)$ is the nonlinear contribution. Under the Bussgang assumptions $\mathbb{E}[\chi_{x,d}(l,0)] = 0$, and
$\mathbb{E}[\chi_{d,x}(l,0)] = 0$.

Moreover, since $\mathbf{t}$ is a unitary transform of the i.i.d. TD distortion $d(n)$, its components $t(n)$ are also independent and identically distributed (\textit{i.e.}, $\mathbb{E}[\mathbf{t}\mathbf{t}^H]=\sigma_t^2 \mathbf{I}$). Therefore, we obtain,
\begin{equation}
\mathbb{E}[\Delta_{\text{Doppler}}] = \mathbb{E}[\chi_{d,d}(0,k)]= \mathbb{E}[\mathbf{t}^H \mathbf{B}_{(0,k)} \mathbf{t}]
=
\sigma_t^2 \operatorname{tr}(\mathbf{B}_{(0,k)}).\label{eq:E[Delta]}
\end{equation}
From \eqref{B_lk}, it is easy to prove that the diagonal terms of $\mathbf{B}_{(0,k)}$ are null for nonzero integer Doppler shifts \cite{yin2025ambiguity}, which implies that in \eqref{eq:E[Delta]}, the contribution of the nonlinear term to the existing sidelobes is zero in expectation. Whereas, the distortion term $\mathbb{E}[\Delta_{\text{Doppler}}]$ contributes exactly $\sigma_t^2 \operatorname{tr}(\mathbf{B}_{(0,0)}) = \sigma_t^2 N$, to the mainlobe. As a result, the Doppler sidelobes decreases after nonlinear amplification, relatively to the mainlobe, which in result decreases the $\mathrm{PSLR}$.

\section{Numerical Results} \label{section3}
We now present numerical results comparing the sensing characteristics of the candidate waveforms, namely AFDM and OFDM. Both $16$-QAM (Quadrature Amplitude Modulation) and $16$-PSK (Phase Shift Keying) constellations are considered to assess the impact of symbol statistics on the ambiguity behavior. The PA is modeled using the Rapp AM/AM characteristic. The parameters are set to $q=1.1$, $G=16$, and $V_{sat}=1.9$ \cite{tani2018papr}, where $q$ denotes the smoothness factor, $G$ the linear gain, and $V_{sat}$ the saturation voltage. The distortion level is controlled through the input back-off (IBO), defined as $\mathrm{IBO} = \frac{V_{sat}}{\sigma_x}$.
All zero-Doppler and zero-delay cuts are evaluated by their normalized average magnitude over 100 Monte Carlo realizations. The chirp frequencies of the $\mathrm{IDAFT}$/$\mathrm{DAFT}$ are appropriately selected to uphold the orthogonality condition \cite{bemani2021afdm} $2(k _\text{max} + \xi)(l_\text{max} + 1) + l_\text{max} \leq N$, where $k_\text{max}$ and $l_\text{max}$ are, respectively, the maximum normalized digital Doppler shift and delay of the channel and $\xi \in \mathbb N_0$ is a free parameter determining the number of additional guard elements around the diagonals to anticipate for Doppler-domain interference.

\subsection{Ambiguity Function Analysis}
\begin{figure}[!t]
  \centering
  \includegraphics[width=1\linewidth]{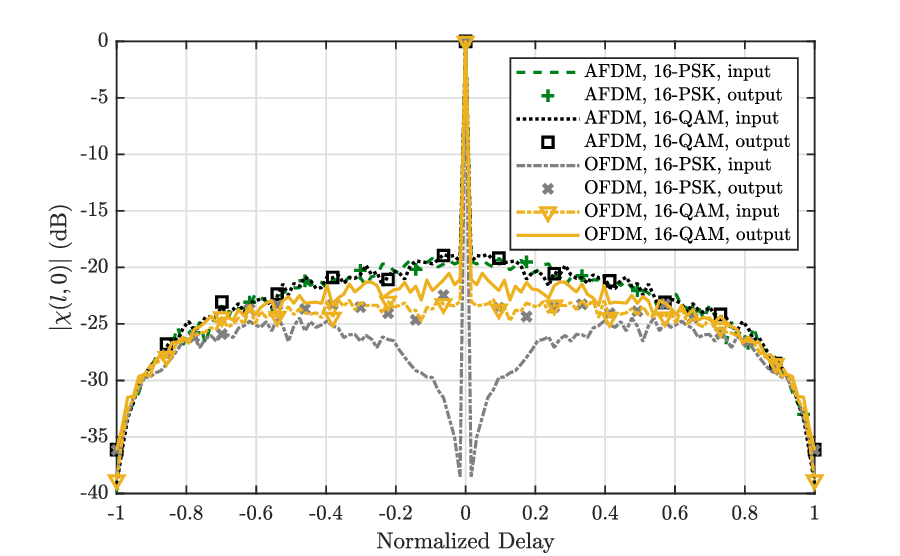}
  \caption{\centering{Zero-Doppler cuts of different constellations under AFDM and OFDM signaling, before and after nonlinear amplification. $N$ = 64, IBO = 1 dB.}}
          \label{fig1}
\end{figure}

Figure~\ref{fig1} represents the zero-Doppler cuts before and after nonlinear amplification for both AFDM and OFDM waveforms using PSK and QAM constellations. A clear increase of the sidelobes is observed for OFDM under strong nonlinear distortion (low IBO). In particular, the sidelobes increase noticeably in the OFDM case, especially with PSK modulation, indicating a higher sensitivity to PA nonlinearities in terms of ranging performance. In contrast, AFDM shows a remarkable insensitivity where the sidelobes remain nearly unchanged before and after amplification. The sidelobe level at the input is initially high, which makes the contribution of the distortion negligible. Furthermore, for AFDM, both PSK and QAM constellations yield almost identical sidelobe levels. This indicates that, under AFDM signaling, the ranging sidelobes are primarily governed by the waveform modulation scheme rather than by the specific constellation geometry.

\begin{figure}[!t]
  \centering
  \includegraphics[width=1\linewidth]{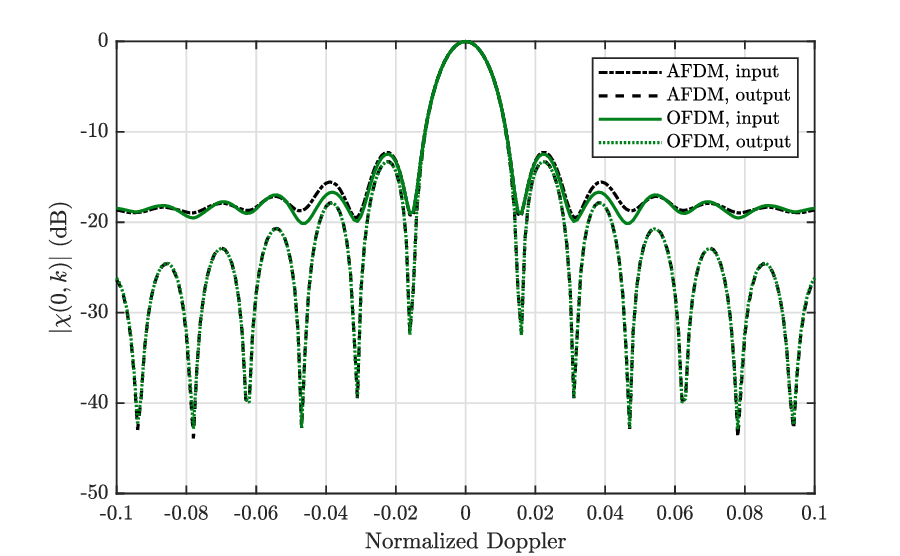}
  \caption{\centering zero-delay cuts of PSK-AFDM \& PSK-OFDM signaling, N = 64.}
          \label{fig3}
\end{figure}

Figure~\ref{fig3} shows the corresponding zero-delay cuts for PSK signaling. Interestingly, both waveforms show comparable Doppler-domain characteristics. Moreover, the amplified signals even display reduced sidelobe levels, consistent with our earlier observations; this reduction can be attributed to compression effects, which smooth out large instantaneous peaks \cite{gourar2025ambiguity}. It should be noted that, despite the significant reduction of the far sidelobes, the sidelobes in the vicinity of the main lobe are slightly reduced.

While qualitative inspection of the ambiguity plots provides useful intuition, a systematic comparison requires quantitative metrics. For this purpose, we evaluate the PSLR, summarized in Table~\ref{tab:pslr_comparison}, for both the delay and Doppler domains. First, in the delay domain, OFDM achieves lower PSLR than AFDM. However, OFDM experiences a noticeable degradation after nonlinear amplification (approximately 2 dB at IBO = 1 dB), whereas AFDM shows only marginal variation (less than 0.5 dB). This confirms that AFDM delay sidelobes are governed by the $\mathrm{IDAFT}$ spreading and are therefore largely unaffected by PA distortions. Second, in the Doppler domain, both waveforms exhibit nearly identical PSLR values, which are reduced after nonlinear amplification.

\begin{figure}[!t]
  \centering
  \subfloat[]{%
  \includegraphics[width=1\linewidth]{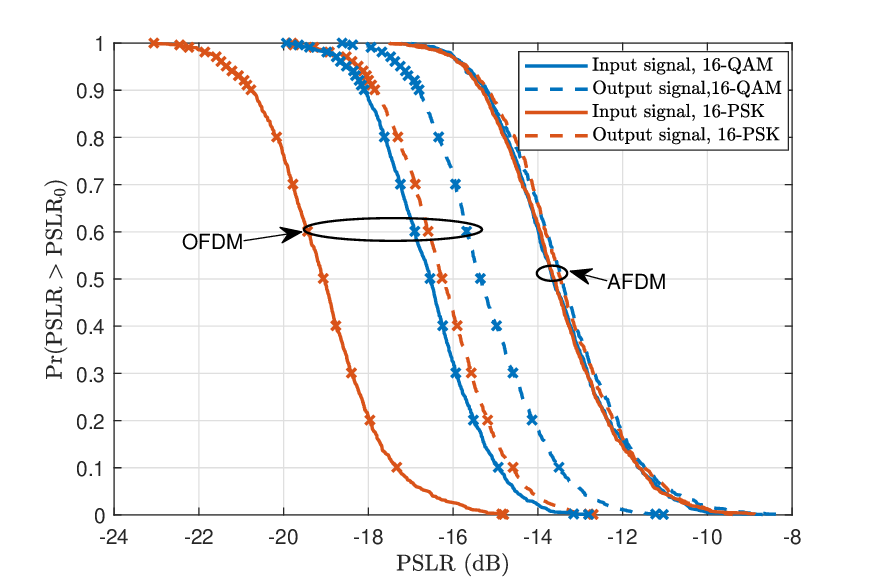}}\\
  \subfloat[]{%
  \includegraphics[width=1\linewidth]{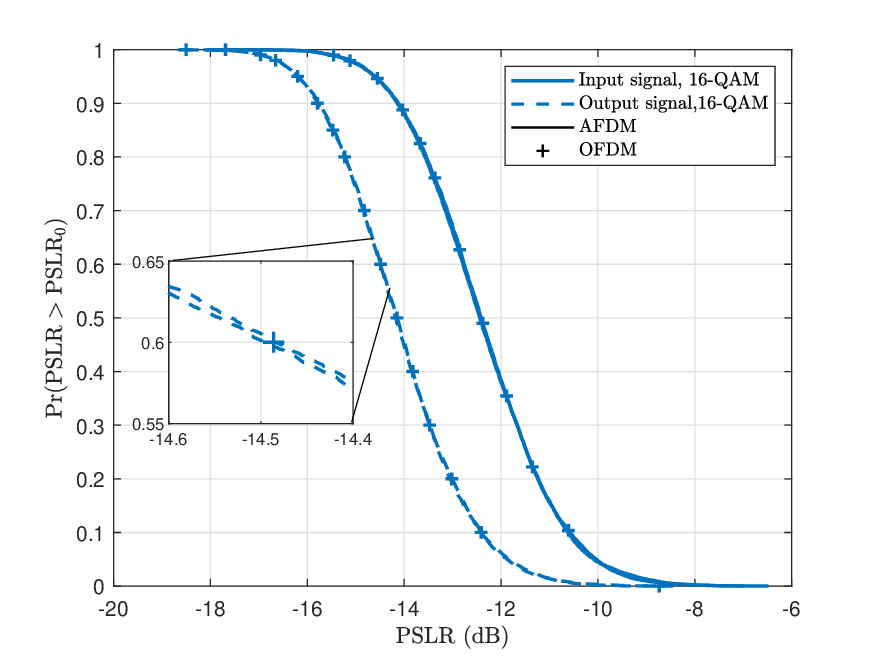}}
    \caption{\centering{CCDF of the $\mathrm{PSLR}$ of the a) zero-Doppler and b) zero-delay cuts, before and after power amplification, IBO = 2 dB.}}
    \label{fig4}
\end{figure}

\begin{table*}[!t]
\centering
\caption{PSLR (dB) Before and After HPA for Delay and Doppler Cuts under 16-QAM and 16-PSK}
\label{tab:pslr_comparison}
\renewcommand{\arraystretch}{1.0}
\setlength{\tabcolsep}{8pt}

\begin{tabular}{llcccccccc}
\toprule
\multirow{2}{*}{\textbf{Modulation}} &
\multirow{2}{*}{\textbf{Waveform}} &
\multicolumn{4}{c}{\textbf{$\mathrm{PSLR}_{\mathrm{dB}}(l,0)$}} &
\multicolumn{4}{c}{\textbf{$\mathrm{PSLR}_{\mathrm{dB}}(0,k)$}} \\
\cmidrule(lr){3-6} \cmidrule(lr){7-10}
 & 
 & \multicolumn{2}{c}{IBO = 1 dB}
 & \multicolumn{2}{c}{IBO = 6 dB}
 & \multicolumn{2}{c}{IBO = 1 dB}
 & \multicolumn{2}{c}{IBO = 6 dB} \\
\cmidrule(lr){3-4} \cmidrule(lr){5-6}
\cmidrule(lr){7-8} \cmidrule(lr){9-10}
 &  & Before & After & Before & After 
 & Before & After & Before & After \\
\midrule

\multirow{2}{*}{16-QAM}
& OFDM  
& -22.62 & -20.64 & -22.77 & -21.01
& -12.30 & -13.27 & -12.70 & -13.30 \\

& AFDM  
& -18.20 & -18.37 & -18.93 & -18.55
& -12.42 & -13.25 & -13.00 & -13.29 \\

\midrule

\multirow{2}{*}{16-PSK}
& OFDM  
& -24.46 & -22.42 & -24.58 & -22.55
& -12.30 & -13.27 & -12.70 & -13.30 \\

& AFDM  
& -18.64 & -18.70 & -18.08 & -18.30
& -12.42 & -13.25 & -13.00 & -13.29 \\

\bottomrule
\end{tabular}
\end{table*}
It is important to note that averaging the AF over many realizations may smooth out rare nonlinearity occurrences and therefore underestimate instantaneous unfavorable ambiguity instances. Motivated by this observation, we present in Fig.~\ref{fig4} the complementary cumulative distribution function (CCDF) of the PSLR for the zero-Doppler and zero-delay cuts, i.e., $\Pr\left(\mathrm{PSLR}>\mathrm{PSLR}_0\right)$, which represents the outage probability that the ambiguity sidelobes exceed a certain threshold $\mathrm{PSLR}_0$. Since larger $\mathrm{PSLR}$ values correspond to higher sidelobes, a left-shifted CCDF indicates improved ambiguity characteristics. Accordingly, for a fixed sidelobe threshold, a lower CCDF value implies a lower outage probability and thus a more robust matched filter response under nonlinearities.
Fig.~\ref{fig4}(a) shows the outage probability associated with the zero-Doppler cut. A clear performance gap between OFDM and AFDM can be observed. Moreover, the AFDM curves undergo only a slight shift after amplification, revealing a significantly smaller increase in outage probability and therefore a lesser sensitivity of the ranging sidelobes to nonlinearities. The outage probability interpretation is particularly meaningful when a target sidelobe level is imposed by the ISAC system design, as it is directly reflected in the range-Doppler response of the matched filter. For instance, near $\mathrm{PSLR}_0=-14$ dB, the OFDM curves exhibit lower outage probabilities than the AFDM curves, meaning that the OFDM sidelobes are less likely to surpass this threshold; however, these probability values are highly sensitive to nonlinear amplification. A different behavior is observed in Fig.~\ref{fig4}(b), corresponding to the zero-delay cut. The output curves are left-shifted, confirming our earlier observations and indicating a smaller outage probability after nonlinear amplification, which is always favorable.

\subsection{Sensing performance}
For the practical detection, we consider the monostatic case where we rely on the knowledge of the transmit signal at the receiver. The carrier frequency $f_c $ is fixed to 24~GHz. For comparison, the OFDM waveform is processed using a matched filter in the time-frequency domain, followed by a 2D RD map construction by the 2D-$\mathrm{FFT}$ \cite{ni2022afdm}.

\begin{figure}[!t]
  \centering
  \includegraphics[width=1\linewidth]{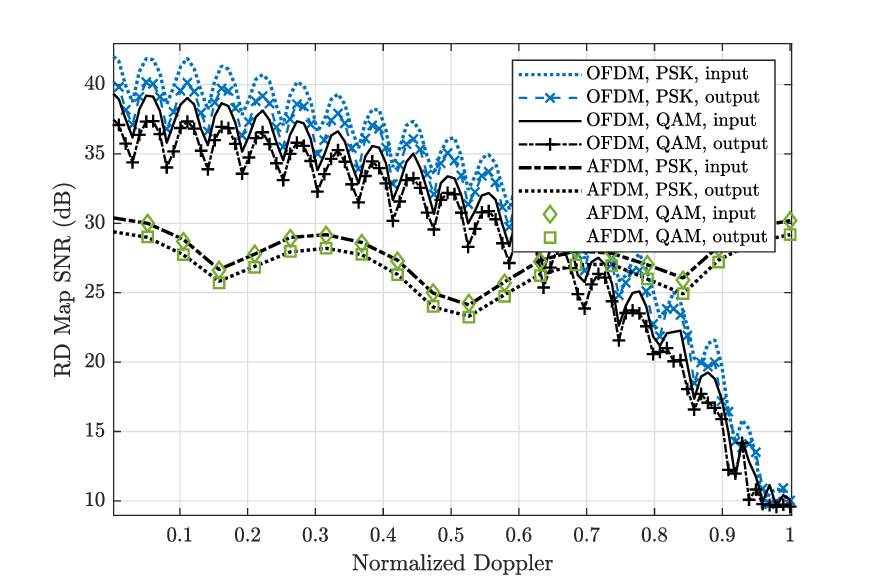}
    \caption{\centering{RD map SNR versus the normalized Doppler shift with $\mathrm{SNR}$=0 dB, $P =1$ and $l=32$}}
    \label{fig5}
\end{figure}

Fig.~\ref{fig5} shows the map SNR, which is the ratio between the peak caused by the target and the average noise level in the RD map, before and after nonlinear amplification, versus normalized Doppler shift (with respect to $\Delta f$ the chirp subcarrier spacing). AFDM exhibits a nearly constant SNR across Doppler shifts, while OFDM shows a pronounced drop near one subcarrier spacing due to inter-carrier interference. Under nonlinear amplification, the radar map SNR degrades with AFDM by approximately 1 dB, compared to about 2 dB for OFDM, and identical curves are observed for PSK and QAM, consistent with the AF cuts' observations.
\begin{figure}[!t]
  \centering
  \includegraphics[width=1\linewidth]{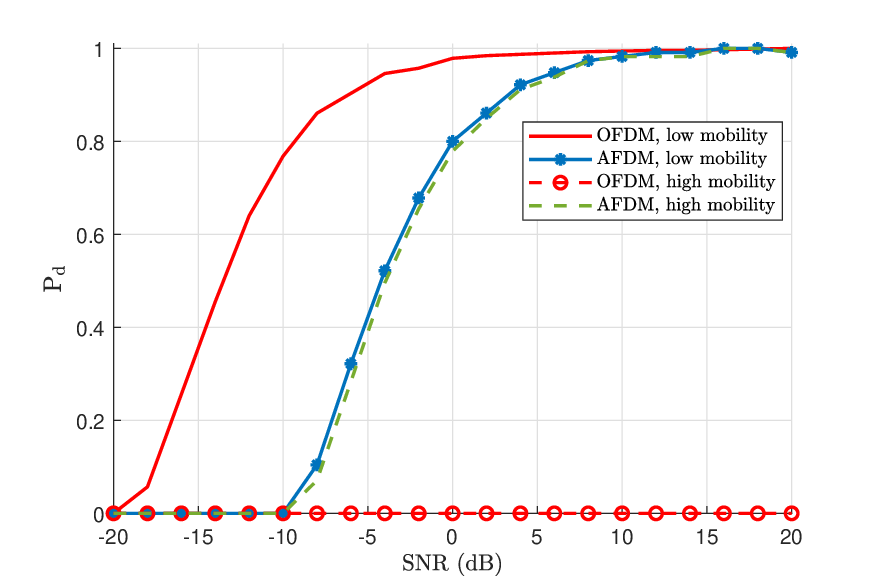}
    \caption{\centering{Probability of detection versus SNR with nonlinearily amplified signals. The solid lines correspond to $f=0.1\Delta f$, and the dashed lines correspond to $f=0.98\Delta f$. $N=1024$ and $M=16$.}}
    \label{fig6}
\end{figure}

Finally, to evaluate the average MF response under PA nonlinearities, we show in Fig.~\ref{fig6} the probability of detection $P_d$ versus SNR for $P=3$ resolvable targets, using 16-PSK symbols. It is computed as the average per-target successfully detected over 1k Monte Carlo realizations. Detection is performed using a CFAR detector \cite{ni2022afdm} over the RD map with a fixed false alarm probability of $10^{-4}$, and SNR is defined as the inverse of the noise variance. At high velocities, OFDM fails to detect the targets. The solid lines correspond to moderate velocities ($f = 0.1\Delta f$), while the dashed lines correspond to high velocities ($f = 0.98\Delta f$). The substantial drop in SNR (as seen in Fig.~\ref{fig5}) due to Doppler-induced inter-carrier interference causes the CFAR detector to trigger on noise peaks rather than the true target peaks. In contrast, AFDM manages to maintain a similar detection probability. As observed in Fig.~\ref{fig5}, the map SNR is constant and remains sufficiently high for the CFAR threshold to detect the targets reliably. Here, the PA introduces an additive distortion term that slightly raises the noise floor, but the peak positions in the RD map remain unbiased, so only minor adjustments to the CFAR threshold are needed.

\section{Conclusion}
We presented a study of the Ambiguity function (AF) and the matched filtering of the AFDM waveform under nonlinear amplification, and compared it to the case of the OFDM waveform. Results show that AFDM is inherently insensitive to Doppler shifts and power amplifier nonlinearities, and its sensing performance is largely independent of the modulation constellation. Constellation shaping offers little room for further AF optimization, suggesting that future improvements should combine other design dimensions, such as chirp-parameter design and pulse shaping or filtering, to further enhance delay-Doppler resolution.

\appendices

\bibliographystyle{IEEEtran}
\bibliography{main.bib}

@inproceedings{tani2018papr,
  title={{PAPR reduction of post-OFDM waveforms contenders for 5G \& Beyond using SLM and TR algorithms}},
  author={Tani, Khaled and Medjahdi, Yahia and Shaiek, Hmaied and Zayani, Rafik and Roviras, Daniel},
  booktitle={2018 25th International Conference on Telecommunications (ICT)},
  pages={104--109},
  year={2018},
  organization={IEEE}
}

@article{liu2025cp,
  title={CP-OFDM achieves the lowest average ranging sidelobe under QAM/PSK constellations},
  author={Liu, Fan and Zhang, Ying and Xiong, Yifeng and Li, Shuangyang and Yuan, Weijie and Gao, Feifei and Jin, Shi and Caire, Giuseppe},
  journal={IEEE Transactions on Information Theory},
  year={2025},
  publisher={IEEE}
}

@inproceedings{ismail2024robustness,
  title={{Robustness of ISAC Waveforms to Power Amplifier Distortion}},
  author={Ismail, Abdur Rahman Mohamed and Guenach, Mamoun and Sakhnini, Adham and Bourdouk, Andr{\'e} and Steendam, Heidi},
  booktitle={2024 IEEE 4th International Symposium on Joint Communications \& Sensing (JC\&S)},
  pages={1--6},
  year={2024},
  organization={IEEE}
}

@article{gourar2025ambiguity,
  title={On the Ambiguity Function of OFDM-based ISAC Signals Under Non-Ideal Power Amplifiers},
  author={Gourar, Eya and Medjahdi, Yahia and Clavier, Laurent and Gizzini, Abdul Karim and Sondi, Patrick},
  journal={arXiv preprint arXiv:2512.09803},
  year={2025}
}

@inproceedings{ni2022afdm,
  title={An AFDM-based integrated sensing and communications},
  author={Ni, Yuanhan and Wang, Zulin and Yuan, Peng and Huang, Qin},
  booktitle={2022 International Symposium on Wireless Communication Systems (ISWCS)},
  pages={1--6},
  year={2022},
  organization={IEEE}
}

@article{rou2024otfs,
  title={From OTFS to AFDM: A comparative study of next-generation waveforms for ISAC in doubly-dispersive channels},
  author={Rou, Hyeon Seok and de Abreu, Giuseppe Thadeu Freitas and Choi, Junil and Kountouris, Marios and Guan, Yong Liang and Gonsa, Osvaldo and others},
  journal={arXiv preprint arXiv:2401.07700},
  year={2024}
}

@article{bemani2023affine,
  title={Affine frequency division multiplexing for next generation wireless communications},
  author={Bemani, Ali and Ksairi, Nassar and Kountouris, Marios},
  journal={IEEE Transactions on Wireless Communications},
  volume={22},
  number={11},
  pages={8214--8229},
  year={2023},
  publisher={IEEE}
}

@inproceedings{bemani2021afdm,
  title={AFDM: A full diversity next generation waveform for high mobility communications},
  author={Bemani, Ali and Ksairi, Nassar and Kountouris, Marios},
  booktitle={2021 IEEE International Conference on Communications Workshops (ICC Workshops)},
  pages={1--6},
  year={2021},
  organization={IEEE}
}

@article{zhu2024afdm,
  title={AFDM-based bistatic integrated sensing and communication in static scatterer environments},
  author={Zhu, Jiajun and Tang, Yanqun and Liu, Fan and Zhang, Xiaoying and Yin, Haoran and Zhou, Yu},
  journal={IEEE Wireless Communications Letters},
  volume={13},
  number={8},
  pages={2245--2249},
  year={2024},
  publisher={IEEE}
}

@article{chafii2023twelve,
  title={Twelve scientific challenges for 6G: Rethinking the foundations of communications theory},
  author={Chafii, Marwa and Bariah, Lina and Muhaidat, Sami and Debbah, Merouane},
  journal={IEEE Communications Surveys \& Tutorials},
  volume={25},
  number={2},
  pages={868--904},
  year={2023},
  publisher={IEEE}
}

@article{zhang2025afdm,
  title={AFDM-enabled integrated sensing and communication: Theoretical framework and pilot design},
  author={Zhang, Fan and Wang, Zhaocheng and Mao, Tianqi and Jiao, Tianyu and Zhuo, Yinxiao and Wen, Miaowen and Xiang, Wei and Chen, Sheng and Karagiannidis, George K},
  journal={IEEE Journal on Selected Areas in Communications},
  year={2025},
  publisher={IEEE}
}

@article{yin2025ambiguity,
  title={Ambiguity function analysis of AFDM signals for integrated sensing and communications},
  author={Yin, Haoran and Tang, Yanqun and Ni, Yuanhan and Wang, Zulin and Chen, Gaojie and Xiong, Jun and Yang, Kai and Kountouris, Marios and Guan, Yong Liang and Zeng, Yong},
  journal={IEEE Journal on Selected Areas in Communications},
  year={2025},
  publisher={IEEE}
}

@article{zhang2025discrete,
  title={On Discrete Ambiguity Functions of Random Communication Waveforms},
  author={Zhang, Ying and Liu, Fan and Xiong, Yifeng and Yuan, Weijie and Li, Shuangyang and Zheng, Le and Han, Tony Xiao and Masouros, Christos and Jin, Shi},
  journal={arXiv preprint arXiv:2512.08352},
  year={2025}
}

@article{bedeer2025ambiguity,
  title={Ambiguity function analysis of affine frequency division multiplexing for integrated sensing and communication},
  author={Bedeer, Ebrahim},
  journal={arXiv preprint arXiv:2504.02582},
  year={2025}
}

@article{rou2025normalized,
  title={Normalized Ambiguity Function Characteristics of OFDM, OTFS, AFDM, and CP-AFDM for ISAC},
  author={Rou, Hyeon Seok and de Abreu, Giuseppe Thadeu Freitas},
  journal={arXiv preprint arXiv:2510.11216},
  year={2025}
}

@article{ni2025ambiguity,
  title={Ambiguity function analysis of AFDM under pulse-shaped random ISAC signaling},
  author={Ni, Yuanhan and Liu, Fan and Yin, Haoran and Tang, Yanqun and Wang, Zulin},
  journal={arXiv preprint arXiv:2511.04200},
  year={2025}
}

@article{bemani2024integrated,
  title={Integrated sensing and communications with affine frequency division multiplexing},
  author={Bemani, Ali and Ksairi, Nassar and Kountouris, Marios},
  journal={IEEE Wireless Communications Letters},
  volume={13},
  number={5},
  pages={1255--1259},
  year={2024},
  publisher={IEEE}
}

@article{dardari2002theoretical,
  title={A theoretical characterization of nonlinear distortion effects in OFDM systems},
  author={Dardari, Davide and Tralli, Velio and Vaccari, Alessandro},
  journal={IEEE transactions on Communications},
  volume={48},
  number={10},
  pages={1755--1764},
  year={2002},
  publisher={IEEE}
}
\end{document}